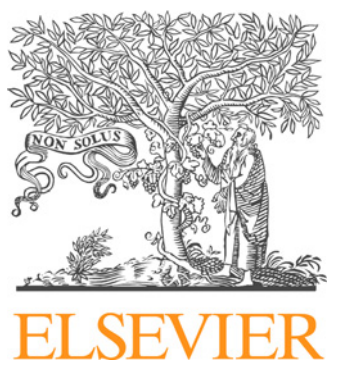

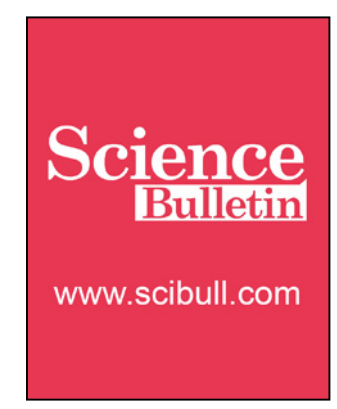

Article

# A persistently active fast radio burst source embedded in an expanding supernova remnant

Chen-Hui Niu [a,1,*], Di Li [b,c,e,1,*], Yuan-Pei Yang [d,1], Yuhao Zhu [c,f,1], Yongkun Zhang [c], Jia-Heng Zhang [a], Zexin Du [a], Jumei Yao [g,h,i], Xiaoping Zheng [a], Pei Wang [c], Yi Feng [j,k], Bing Zhang [l], Weiwei Zhu [c], Wenfei Yu [m], Ji-An Jiang [n,s], Shi Dai [o,p], Chao-Wei Tsai [c,f], A. Ming Chen [a,q,r], Yijun Hou [c], Jiarui Niu [c], Weiyang Wang [f], Chenchen Miao [e], Xinming Li [a], Junshuo Zhang [c,f]

[a] Institute of Astrophysics, Central China Normal University, Wuhan 430079, China
[b] New Cornerstone Science Laboratory, Department of Astronomy, Tsinghua University, Beijing 100084, China
[c] National Astronomical Observatories, Chinese Academy of Sciences, Beijing 100101, China
[d] South-Western Institute for Astronomy Research, Yunnan University, Kunming 650504, China
[e] Research Center for Astronomical Computing, Zhejiang Laboratory, Hangzhou 311100, China
[f] School of Astronomy and Space Science, University of Chinese Academy of Sciences, Beijing 100049, China
[g] Xinjiang Astronomical Observatory, Chinese Academy of Sciences, Urumqi 830011, China
[h] Key Laboratory of Radio Astronomy, Chinese Academy of Sciences, Urumqi 830011, China
[i] Xinjiang Key Laboratory of Radio Astrophysics, Urumqi 830011, China
[j] Research Center for Astronomical Computing, Zhejiang Laboratory, Hangzhou 311100, China
[k] Institute for Astronomy, School of Physics, Zhejiang University, Hangzhou 310027, China
[l] Department of Physics and Astronomy, University of Nevada, Las Vegas, LV 89154, USA
[m] Shanghai Astronomical Observatory, Chinese Academy of Sciences, Shanghai 200030, China
[n] Department of Astronomy, University of Science and Technology of China, Hefei 230026, China
[o] Australia Telescope National Facility, CSIRO, Space and Astronomy, Locked Bag, PO Box 76, Epping, NSW 1710, Australia
[p] Western Sydney University, Locked Bag, Penrith South DC, NSW 2751, Australia
[q] Tsung-Dao Lee Institute, Shanghai Jiao Tong University, Shanghai 201210, China
[r] Key Laboratory of Quark and Lepton Physics (MOE), Central China Normal University, Wuhan 430079, China
[s] National Astronomical Observatory of Japan, Tokyo 181-8588, Japan

ABSTRACT

Fast radio bursts (FRBs) remain one of the most puzzling astrophysical phenomena. While most FRBs are detected only once or sporadically, we present the identification of FRB 20190520B as the first persistently active source over a continuous span of ∼4 years. This rare long-term activity enabled a detailed investigation of its dispersion measure (DM) evolution. We also report that FRB 20190520B exhibits a substantial decrease in DM at a global rate of $(-12.4 \pm 0.3)$ pc cm$^{-3}$ yr$^{-1}$, exceeding previous FRB's DM variation measurements by a factor of three and surpassing those observed in pulsars by orders of magnitude. The magnitude and consistency of the DM evolution, along with a high host DM contribution, strongly indicate that the source resides in a dense, expanding ionized medium, likely a young supernova remnant (SNR).

## 1. Introduction

Since the first FRB, also known as the Lorimer burst [1], has been discovered, the extragalactic DM budget has always been one of the characteristics of FRBs. The definition of DM is the integral of ionized media along the LOS, which can be expressed as $\mathrm{DM} = \int_0^D n_e dl$, where $n_e$ is the column density of electrons and $D$ is the distance from the source to the observer. The FRBs DM budget can be used as a tool to measure the missing baryon if the estimated DM contribution from the host galaxy is estimated [2]. FRB 20190520B was first detected by the Five-hundred-meter Aperture Spherical radio Telescope (FAST [3]) in 2019 and then localized to a dwarf galaxy at redshift of ∼0.241 [4]. As an outlier from the $z$-$\mathrm{DM_E}$ relation [2], the unusually high $DM_{host}$ suggests a highly complex environment within the host galaxy [4–6]. The complexity of its environment is also reflected in the detection of rare circular polarization and a highly stochastic, Brownian motion-like bursting behavior in the time-energy domain [7,8].

* Corresponding authors.
*E-mail addresses:* niuchenhui@ccnu.edu.cn (C.-H. Niu), dili@tsinghua.edu.cn (D. Li).
[1] These authors contributed equally to this work.

Variations in DM suggest reconfigurations in the line-of-sight (LOS) ionized plasma density, with significant DM evolution requiring a dense and dynamic ionized environment. Temporal DM variations were first detected in the Crab pulsar PSR B0531 + 21 [9], and subsequent studies have revealed similar variations in some other pulsars. For pulsar observations, the ΔDM varies on a yearly timescale within a range of $\sim 10^{-4}$ to $10^{-3}$ pc cm$^{-3}$ [10–13]. These variations are explained as the parallel or transverse pulsar motion in fluctuated electron density region [14–17]. The ΔDM of binary pulsars can increase significantly near periastron, as exemplified by the well-known B1259-63, whose ΔDM can reach 10–20 pc cm$^{-3}$ [18,19]. This enhanced DM diminishes during the eclipse phase, with variations spanning for ~10 days. Some pulsars show a linear or periodic change in the DM over long observation [12,20–24].

FRBs often exhibit significant variations in DM, even on a burst-to-burst basis. For example, FRB 20201124A, which inhabits a low-stellar-density environment, displays a modest DM variation of at most $|\Delta DM| \leqslant 2.6$ pc cm$^{-3}$ in one year [25]. Such variations may primarily arise from uncertainties in single-burst DM measurements on millisecond timescales for FRBs rather than integrated profiles averaged for pulsars. Similar to pulsars, linear trends in DM variations have also been identified in FRB sources. FRB 20121102A has a positive DM gradient of (+0.85 ± 0.10) pc cm$^{-3}$ yr$^{-1}$ over a 9-year span [26], while FRB 20180301A has a linear decreasing trend at a rate of (–2.7 ± 0.2) pc cm$^{-3}$ yr$^{-1}$ [27].

In this work, we present ~4 year's follow-up observations of FRB 20190520B with FAST. Notably, unlike other active repeating FRBs with shutdown windows, we were able to detect the bursts from the FRB 20190520B in every FAST observation, which is facilitating detailed propagation analysis. The average detection rate of FRB 20190520B by FAST is $(5.6 \pm 5.5)$ h$^{-1}$, and the sustained activity of this source allows us to track fluctuations in the propagation effect and burst rate over time. Consistent with the model of Ref. [28], our findings suggest that FRB 20190520B originates from a young magnetar embedded in an expanding SNR, whose age is inferred to be ~10–100 years based on the observed DM decrease.

## 2. Data

The observation windows for FRB 20190520B were irregularly and randomly assigned by the FAST scheduling team. The observations were carried out using the central feed and pulsar backend system operating in the L band (1.0–1.5 GHz) with a center frequency of 1.25 GHz [29]. The data have a time resolution of 49.152 μs and a frequency resolution of 0.122 MHz, and were recorded in the standard PSRFITS format [30]. Burst searches were performed using the HEIMDALL pipeline, following the procedures described in Refs. [4,31,32]. A zero-DM matched filter was applied for RFI mitigation during the blind search. Candidates with signal-to-noise ratios (S/N) > 7 were retained and visually inspected through their dynamic spectra.

The previous discovery paper on FRB 20190520B reported 75 follow-up bursts in FAST tracking mode between April 25th and September 19th, 2020 [4]. Subsequent FAST observations, spanning October 24th, 2020, to April 1st, 2023, accumulated a total of 86.94 h in 87 observations. In total, 360 new bursts were identified and the event rates spanned from 1 to 27 bursts per hour. Although the event rate of FRB 20190520B is lower than that of FRB 20121102 or FRB 20201124A ($\mathcal{O}(10^2)$ h$^{-1}$) [25,26], the long-term activity provides a unique opportunity to study the evolution of propagation effects. To ensure measurement consistency, we measure the structure-optimized DMs of 435 bursts from all FAST observations, which includes re-measuring 75 single bursts from previous studies. Flux calibration was performed using injected periodic signals with a reference noise temperature before each observation session. Burst energetics were determined through measurements of peak flux density and specific fluence ($F_\nu$) from calibrated data. We measured burst widths and frequency bandwidth using the full width at half maximum (FWHM) from Gaussian fits to the temporal profiles and its bandpass, respectively. The isotropic equivalent burst energy was calculated using method described in Refs. [4,33,34].

## 3. Dispersion measure evolution

We optimized burst dispersion measures using DM_PHASE software,[2] which maximizes coherent signal power rather than S/N, preserving intrinsic temporal structures [35]. This approach, robust against complex morphologies and interference, was applied iteratively across ranges of ±20, ±10, and ±5 pc cm$^{-3}$ with 0.01 pc cm$^{-3}$ steps from initial values. Visual inspection of dedispersed dynamic spectra confirmed no significant anomalies.

The substantial DM variation over the 4 years is evident, as shown in the top panel of Fig. 1, which clearly illustrates the changes. An energy-weighted daily average of DM values was employed to minimize the effect of measurement uncertainties caused by faint bursts. The DM results from all bursts were fitted with a linear model to determine the DM decreasing rate. The derived linear slope corresponds to a DM rate of $(-12.4 \pm 0.3)$ pc cm$^{-3}$ yr$^{-1}$, with an absolute change surpassing other FRBs by a factor of four and many orders of magnitude exceeding the variation rate for pulsars of $\sim 10^{-3}$ pc cm$^{-3}$ yr$^{-1}$ [24]. The measured DM variations for each burst are shown in the Fig. S1 (online).

The structure function (SF) of DM, $D_{\mathrm{DM}}(\tau)$, reveals the small and large time scale variation of DM value in the following:

$$D_{\mathrm{DM}}(\tau) = \left\langle [\mathrm{DM}(t) - \mathrm{DM}(t+\tau)]^2 \right\rangle. \quad (1)$$

In pulsar observations, $D_{\mathrm{DM}}(\tau)$ is used for diffractive scintillation time scale analysis [36,37]. The structure function of turbulence comprises three regimes: noise at short lags, power-law scaling (index ~5/3 for Kolmogorov turbulence) at intermediate lags, and saturation, marked by plateauing or oscillations, at long lags. As illustrated in Fig. 2, the large time scale ($\tau \gtrsim 350$ days) DM SF of FRB 20190520B is well power-law fitted, meanwhile the small scale ($\tau \lesssim 100$ days) is flat noise-dominated. The $D_{\Delta \mathrm{DM}}$, subtracting the linear DM trend, shows no linear trend but in the noise-dominated region, suggests that the large-scale linear trend in $D_{\mathrm{DM}}(\tau)$ is driven by the corresponding DM variation trend. The $D_{\mathrm{DM}}(\tau)$, derived from the initial DM values, yields a power-law index of $2.08 \pm 0.19$, consistent with a linear DM variation. The fitted DM structure function further supports a physical interpretation involving a large-scale, orderly process, possibly associated with an expanding supernova remnant.

Fig. 3 further visualizes the dynamic spectrum of de-dispersed bursts across different DM values. The left and right panels of Fig. 3 correspond to bursts 95 (B95) and 430 (B430). The B430 dynamic spectrum overlays the dedispersed spectra of DM 1200.9 pc cm$^{-3}$ (from B95, recorded 27 months earlier) and the optimal DM 1170.8 pc cm$^{-3}$. When DM 1200.9 pc cm$^{-3}$ is applied to B430, a significant tilt is observed in the dynamic spectrum. The top-right panel compares the burst profiles, revealing a significant S/N reduction when the B95 DM is used.

[2] https://github.com/danielemichilli/DM_phase.

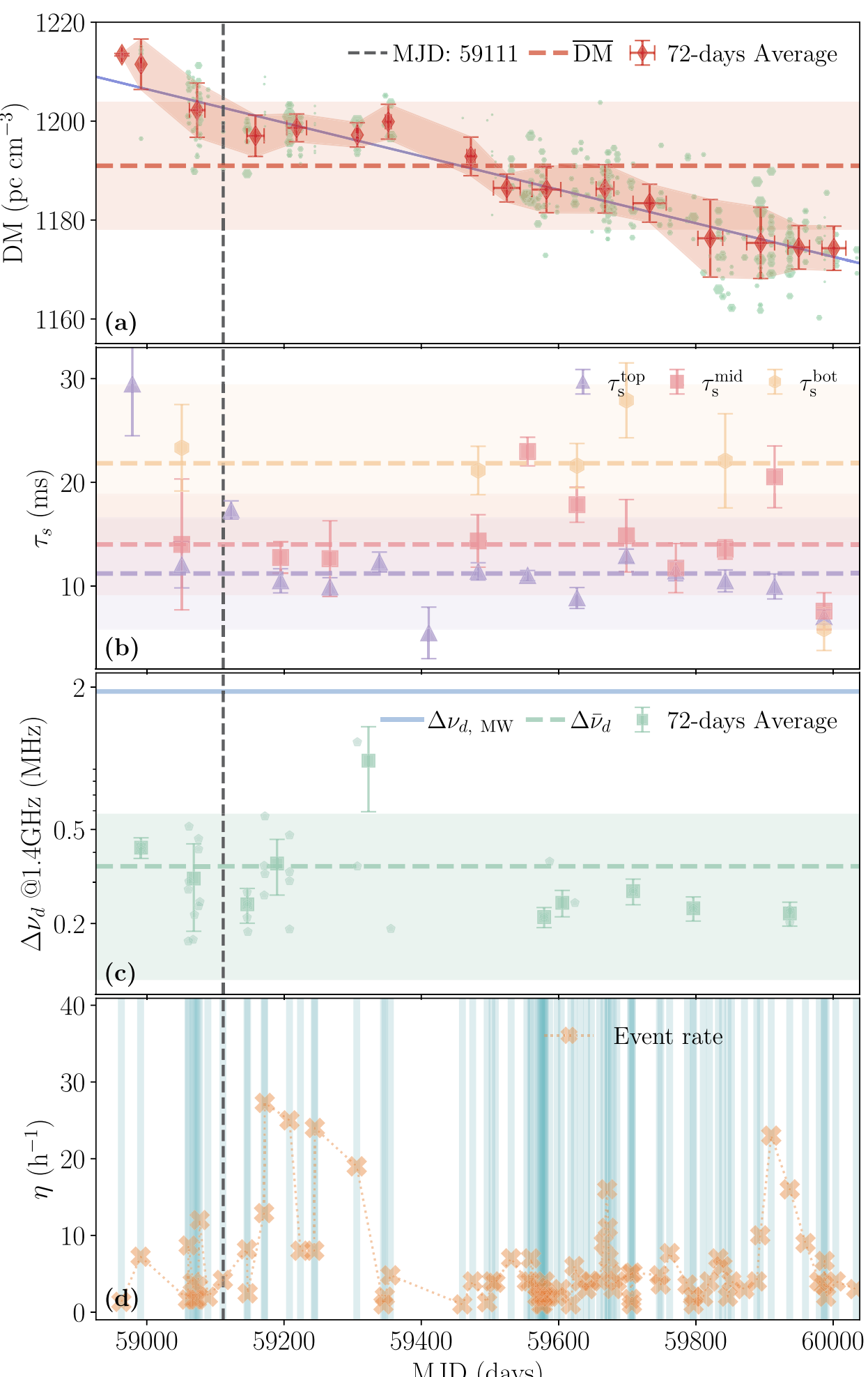

**Fig. 1.** The evolution of the propagation effect for FRB 20190520B. The top 3 panels (a), (b) and (c) depict the evolution of DM, scattering timescale, and scintillation bandwidth. The horizontal dashed lines and shaded areas in these panels denote the mean values and 3$\sigma$ intervals of the measured data. The black dashed vertical line at 59111.37098 marks the boundary, with data before this point coming from previous studies. The red diamonds in panel (a) represent 72-day averaged DM values, and the green hexagram denote the DM values for individual burst, with marker size scaled by burst energy. Triangles, squares, and hexagons in panel (b) represent $\tau_s$ measured in 72-day averaged on the top (1300–1500 MHz), middle (1150–1300 MHz), and bottom (1000–1150 MHz) bandpass, respectively. The blue vertical solid line in panel (c) indicates the scintillation bandwidth resulting from the Milky Way's DM contribution. Panel (d) shows how the event rate varies with date revealing this source's long-term activity. The vertical cyan regions represent the total observation windows.

## 4. Turbulent media propagation

Multi-path propagation induces scattering and scintillation, which are commonly seen in both pulsar and FRB observations [38]. The scattering timescales inferred from burst profiles trace density fluctuations along the LOS. The mean scattering timescale $\tau_s$ of FRB 20190520B is $\sim$10 ms at 1 GHz, but changes rapidly over timescales as fast as minutes [39]. These scattering variations are orders of magnitude faster than variations typical for the ISM, and are most likely from the circumstance medium of FRB 20190520B [39].

The bursts from FRB 20190520B were emitted at different central frequencies. As shown in Fig. 4, FAST observations show that 84% of the bursts occur in the 1300–1500 MHz band, while only 4% are detected in the 1000–1150 MHz band. The results indicate that FRB 20190520B tends to emit more bursts at higher frequencies, which is consistent with the analysis from previous work [40]. In this work, the averaged scattering timescale was measured across three sub-bands in the Fourier spectrum: 1000–1150, 1150–1300, and 1300–1500 MHz.

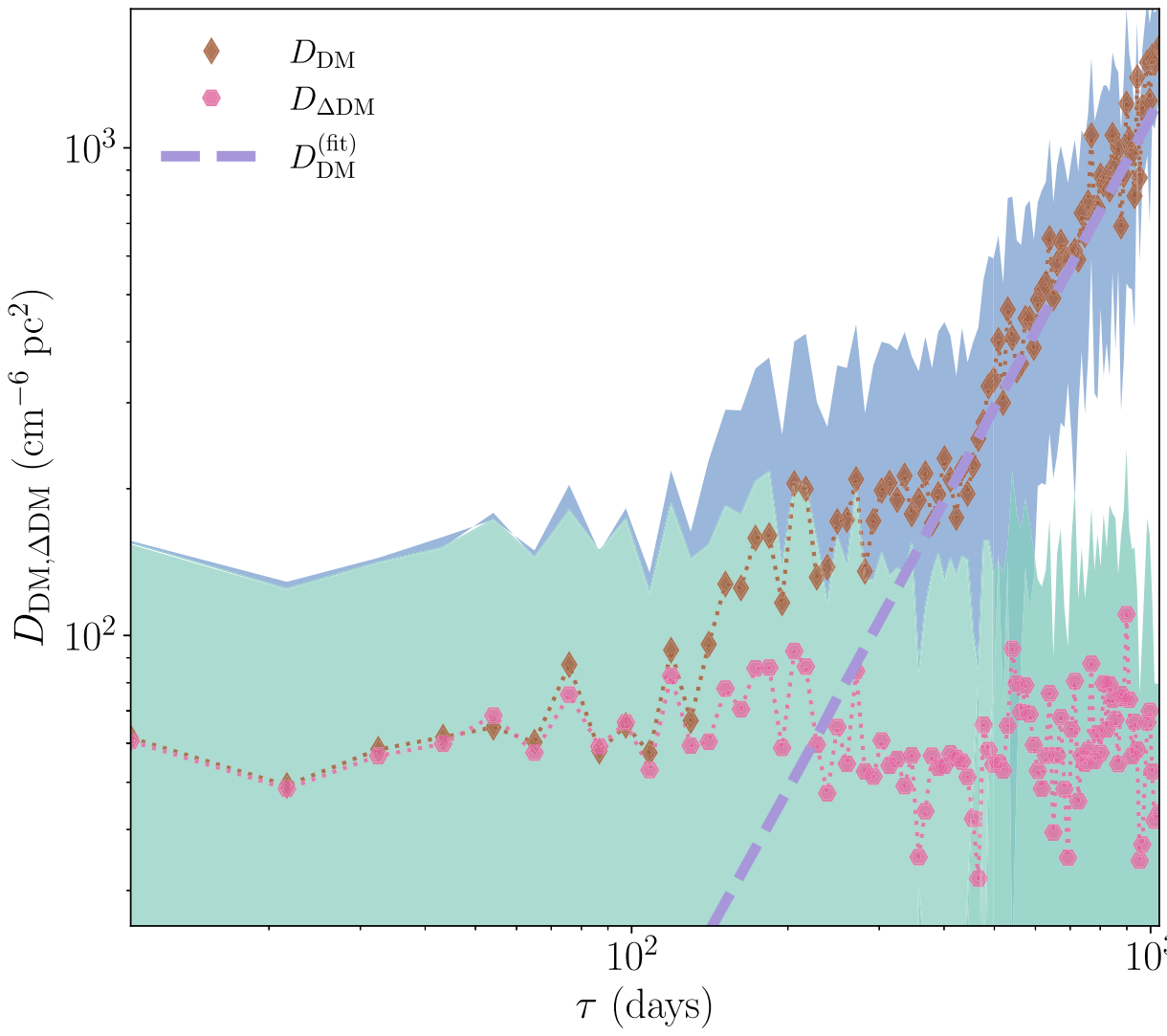

**Fig. 2.** The structure function of DMs for FRB 20190520B. The structure function $D_{\mathrm{DM}}(\tau)$ is derived from the DM values of all the 435 detected bursts. The brown diamonds ($D_{\mathrm{DM}}$) represent the structure function calculated from the measured DM values. At large time lags, the SF exhibits a steep rise that follows the purple dashed line, corresponding to a power-law fit with an index of 2.08 $\pm$ 0.19, consistent with a dominant linear trend. The pink hexagons ($D_{\Delta\mathrm{DM}}$) show the SF of the de-trended data, obtained after subtracting the best-fit linear model. This de-trended SF remains flat, indicating no significant turbulence signal above the noise level. The shaded regions denote 1$\sigma$ uncertainties.

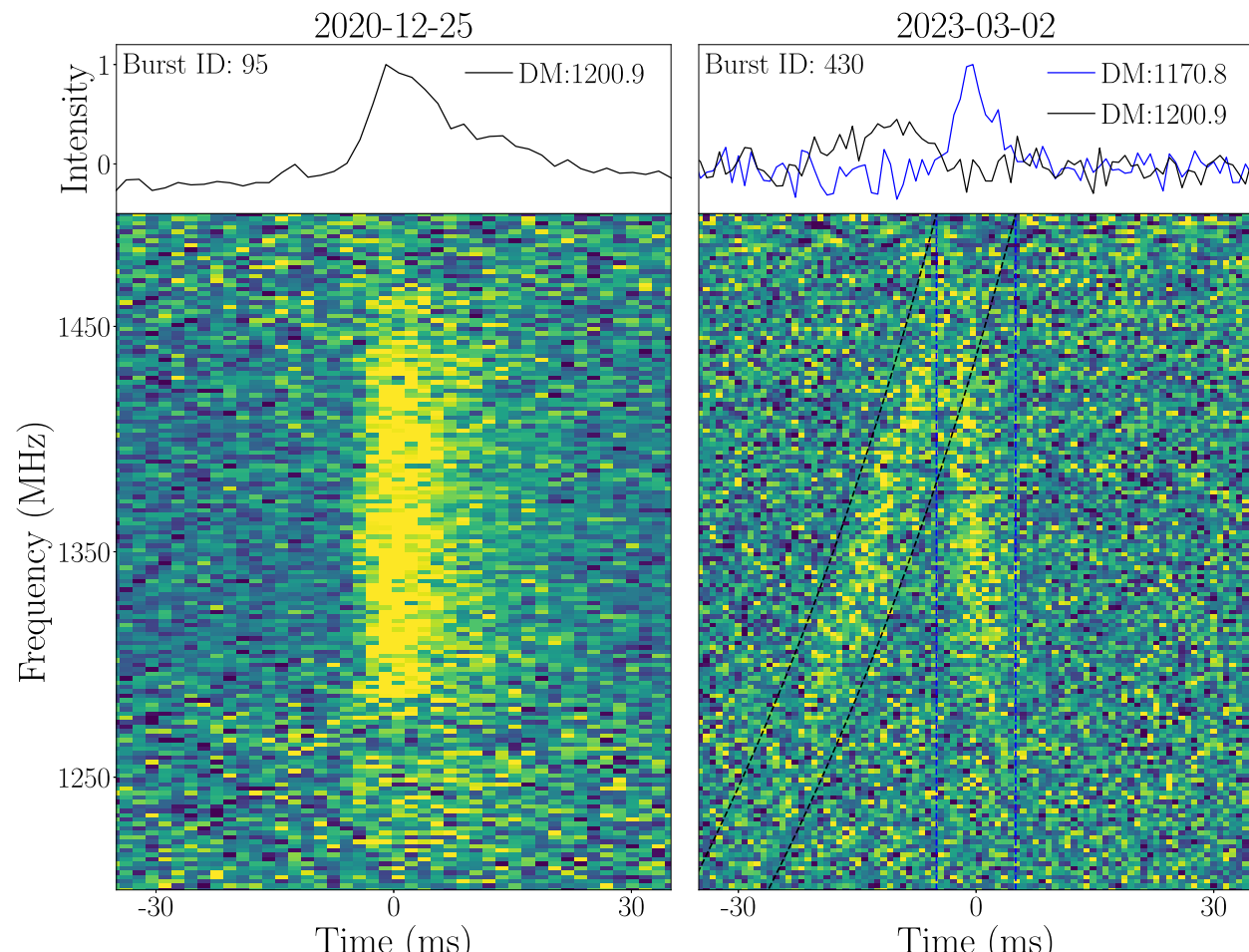

**Fig. 3.** The waterfall plots correspond to yearly duration. The top panel displays the burst profile, while the bottom panel shows the corresponding waterfall plot. The left plots present data from the 95th burst observed on 2020-12-25, whereas the right panel depicts data from the 430th burst observed on 2023-02-02. In the top panel, the black line represents the profile dedispersed with a DM of 1200.9 pc cm$^{-3}$, and the blue line corresponds to the profile dedispersed with a DM of 1170.8 pc cm$^{-3}$, the measured value for the 430th burst. The right waterfall figure combines dedispersed data with DMs of 1200.9 and 1170.8 pc cm$^{-3}$, highlighting the over-dedispersion effect at 1200.9 pc cm$^{-3}$.

Applying the Fourier domain methodology from Refs. [39,41] to the expanded dataset of FAST follow-up observations yields measured $\tau_s$ values ranging widely from 3.3 to 24 ms, with an average of (11 $\pm$ 6) ms, affirming that the scattering time continues to vary significantly. The sub-band-dependent scattering timescales are displayed in Fig. 5 and indicate a frequency-dependent broadening consistent with previous studies. The day-averaged $\tau_s$ varied with time, but no linear trend, as seen in the DM measurements, was identified. Excluding bursts with complex or multi-component structures, we performed a stacking analysis of the remaining profile spectra.

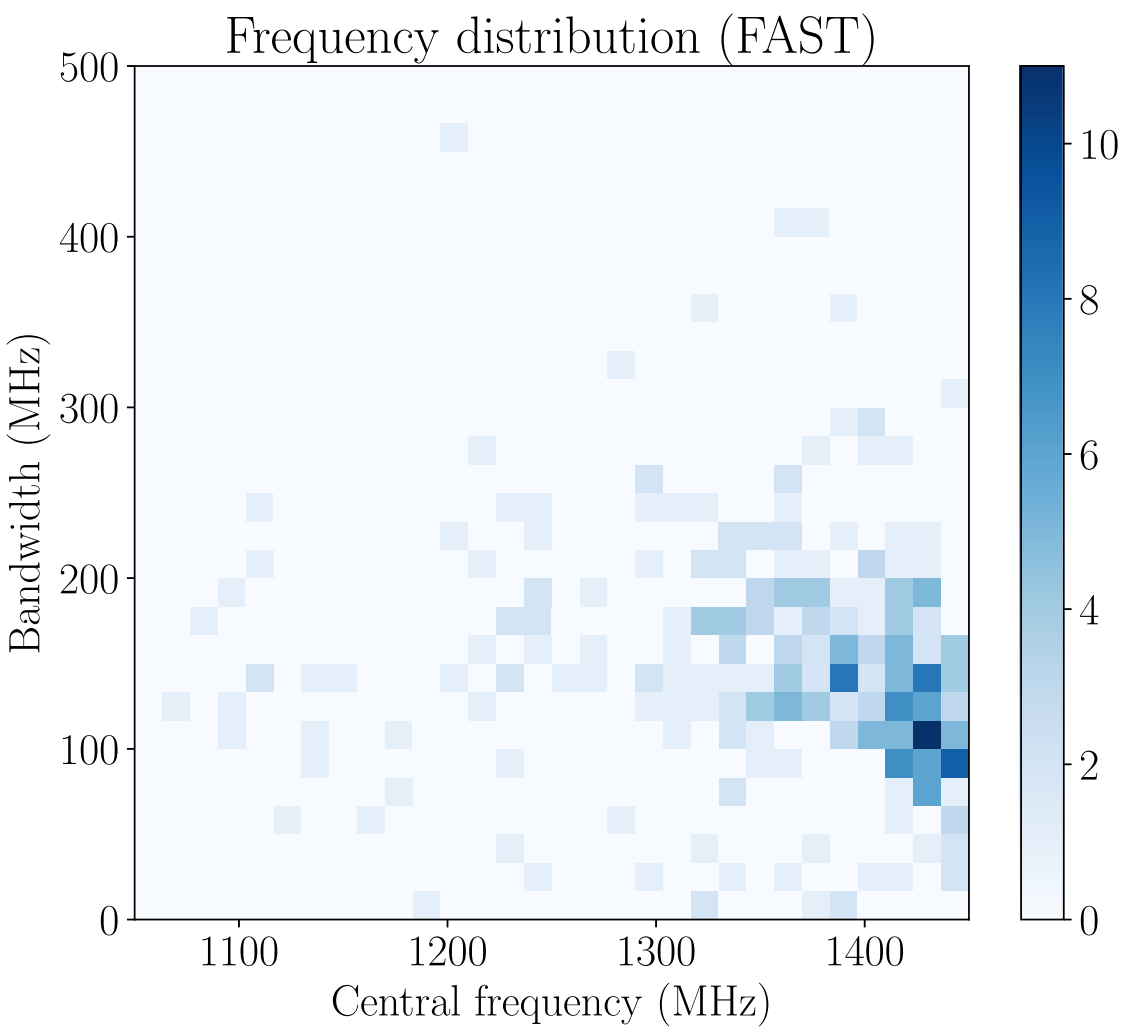

**Fig. 4.** The distribution of central frequencies and bandwidth for FRB 20190520B. The central frequencies of most bursts from FRB 20190520B are concentrated in the higher end of FAST's frequency band, suggesting that the intrinsic central frequencies of these events may lie at relatively higher frequencies. The bandwidths of individual bursts also exhibit significant diversity.

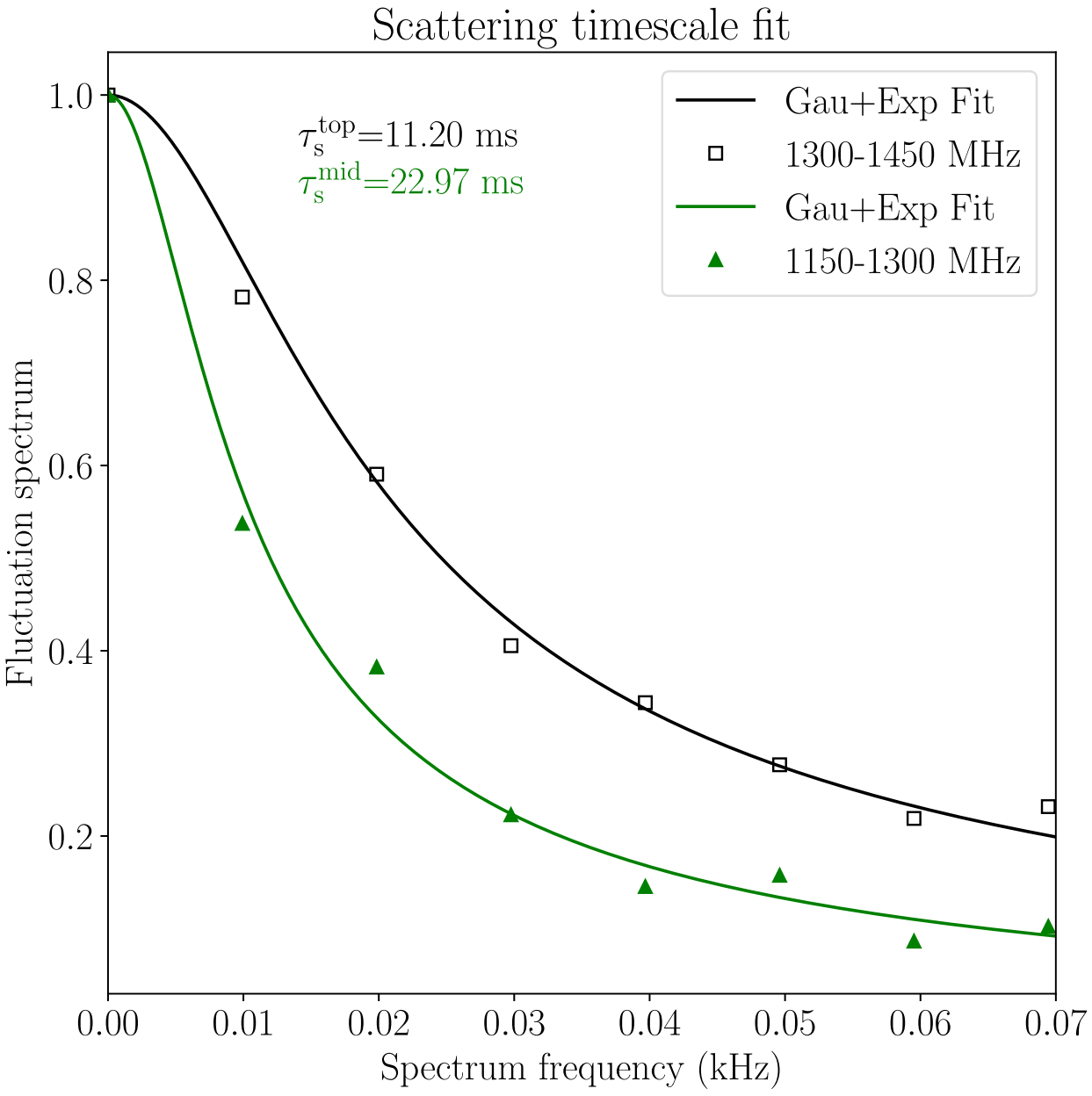

**Fig. 5.** The scattering timescale, $\tau_s$, was obtained by fitting the stacked profile spectrum of bursts from FRB 20190520B. Following the method proposed by Refs. [39,41], we analyzed burst profiles in the Fourier domain across three sub-bands and performed fits to derive the scattering timescales, $\tau_s$.

We additionally measured scintillation bandwidths for 25 out of the 435 bursts, selecting those whose autocorrelation function (ACF) had a signal-to-noise ratio (S/N) exceeding 7. The median $\nu_d$ at 1.4 GHz, $(0.27 \pm 0.12)$ MHz, is consistent with the value expected for the turbulent ISM in the Milky Way and is in agreement with the earlier result of $(0.21 \pm 0.01)$ MHz at 1.4 GHz [39]. These results support the previous two-screen interpretation that the scintillation comes from the Milky Way and the extragalactic scattering from near the source.

To evaluate cross-correlation and the Pearson coefficient $(\rho)$, we interpolated the measured parameters as a function of time. Based on the association between the scattering time variations and circumstance medium, we searched for correlations between the variations in $\tau_s$ and DM. No significant correlation between $\tau_s$ and DM was found.

## 5. Results

The DM evolution of FRB 20190520B shows an extremely large host DM of DM $\simeq 902$ pc cm$^{-3}$ and a long-term decline in the past few years with a decline rate of $\mathrm{dDM/d}t = (-12.4 \pm 0.3)$ pc cm$^{-3}$ yr$^{-1}$. Lee et al. [6] recently found that two separate galaxy clusters are directly intersected by the FRB 20190520B sightline. After subtracting off their DM contributions from galaxy halo, the host DM is estimated to be 280 – 430 pc cm$^{-3}$. Although such a result implies that the host DM of FRB 20190520B might no longer significantly exceeds the predicted values estimated by DM-$z$ relation, the observed long-term DM decline reported by this work supports that there is still a significant contribution from the nearby plasma, because galaxy clusters can not contribute an observable DM variation [42]. Consider the uncertainty of the DM contribution from the foreground galaxy clusters, in the following discussion, we will discuss a possible range of the host DM from 300–900 pc cm$^{-3}$. The long-term monotonic decrease in DM, together with the significant host contribution of FRB 20190520B, is consistent with the expected signature of an expanding shell such as a supernova remnant, as suggested by previous work [42,43]. During the SNR expansion, both the electron density and the SNR thickness evolve, leading to a power-law evolution, DM $\propto t^{-\alpha}$, where the temporal index $\alpha$ depends on the evolution stages of the SNR and the distribution of the ionized medium in the SNR, see the detailed discussions in Refs. [42,43]. For example, for a fully ionized ejecta in the free-expansion phase, one has $\alpha = 2$; for a shocked-ionized ejecta in the free-expansion phase, one has $\alpha = 1/2$; for the Sedov-Taylor phase with an ionization degree less than 40%, one has $\alpha = -2/5$. According to the above DM power-law evolution, the SNR age could be generally written as

$$t_{\rm SNR} \simeq \left|\frac{\alpha \mathrm{DM}}{\mathrm{dDM/d}t}\right|. \tag{2}$$

The host DM of FRB 20190520B might be

$$\mathrm{DM_h} \simeq (300 - 900)\ \mathrm{pc\ cm^{-3}},$$

as pointed out before [4,6], and the DM decline rate in the long term is $\mathrm{dDM/d}t = (-12.4 \pm 0.3)$ pc cm$^{-3}$ yr$^{-1}$ based on this work. Assuming that the SNR mainly dominates the host DM, the SNR age is estimated to be $t_{\rm SNR} \simeq (24 - 73)\alpha$ yr. Since the temporal index $\alpha$ is the order of magnitude of the unity as pointed out above, the above result suggests that the SNR should have an age of $\sim 10 - 100$ years, most likely in the free-expansion stage. For a fully ionized ejecta and a shocked-ionized ejecta with $\alpha = 2, 1/2$, respectively, the SNR age is estimated to be $t_{\rm SNR} \sim 50 - 150$, $12 - 37$ years, respectively. One can further estimate the evolution of dDM/dt, i.e., $\mathrm{d^2DM/d}t^2$. We consider that

$$\mathrm{dDM/d}t = -12.4\ \mathrm{pc\ cm^{-3}\ yr^{-1}} (t/t_{\rm SNR})^{-\alpha} \tag{3}$$

with $t_{\rm SNR} \sim 100$ yr. Then one has

$$\mathrm{d^2DM/d}t^2 = (12.4\alpha/t_{\rm SNR})\mathrm{pc\ cm^{-3}\ yr^{-2}} (t/t_{\rm SNR})^{-\alpha-1}. \tag{4}$$

Since the factor of $12.4\alpha/t_{\rm SNR}$ is the order of $\sim 0.1$pc cm$^{-3}$ yr$^{-2}$, it is not easy to measure $\mathrm{d^2DM/d}t^2$ with such a small value.

Based on the above SNR's age, the typical parameters (e.g., the ejecta mass $M$, the ejecta kinetic energy $E$, etc.) of the SNR could be further constrained. For the fully ionized ejecta in the free-expansion phase, the time-dependent DM is [42]

$$\mathrm{DM} \simeq 260\ \mathrm{pc\ cm^{-3}} \left(\frac{M}{10M_\odot}\right)^2 \left(\frac{E}{10^{51}\,\mathrm{erg}}\right)^{-1} \times \left(\frac{t_{\mathrm{SNR}}}{100\mathrm{yr}}\right)^{-2}. \tag{5}$$

Taking $\mathrm{DM_h} \simeq (300-900)\ \mathrm{pc\ cm^{-3}}$ and $t_{\mathrm{SNR}} \simeq (50-150)\ \mathrm{yr}$ for $\alpha = 2$, one finally obtains the following constraint:

$$\left(\frac{M}{10M_\odot}\right)^2 \left(\frac{E}{10^{51}\ \mathrm{erg}}\right)^{-1} \simeq 0.3-7.8. \tag{6}$$

The SNR velocity is given by

$$v \simeq \left(\frac{2E}{M}\right)^{1/2} \simeq (1900-4300)\ \mathrm{km\ s^{-1}} \left(\frac{E}{10^{51}\ \mathrm{erg}}\right)^{1/4}. \tag{7}$$

According to the estimated SNR age of $t_{\mathrm{SNR}} \simeq (50-150)$ yr, the SNR radius is estimated to be

$$R_{\mathrm{SNR}} \simeq v t_{\mathrm{SNR}} \simeq 0.2\mathrm{pc} \left(\frac{E}{10^{51}\ \mathrm{erg}}\right)^{1/4}. \tag{8}$$

For the shocked-ionized ejecta in the free-expansion phase, the time-dependent DM is [43]

$$\mathrm{DM} \simeq 29.5\,\mathrm{pc\,cm^{-3}} \left(\frac{M}{10M_\odot}\right)^{3/4} \left(\frac{E}{10^{51}\,\mathrm{erg}}\right)^{-1/4} \times \left(\frac{n_0}{1\ \mathrm{cm^{-3}}}\right)^{1/2} \left(\frac{t_{\mathrm{SNR}}}{100\ \mathrm{yr}}\right)^{-1/2}, \tag{9}$$

where $n_0$ is the number density of a uniform ambient ISM. Here, the mean molecular weight $\mu$ and the mean molecular weight per electron $\mu_e$ are potentially assumed to be approximately equal, $\mu \sim \mu_e$, see Ref. [43]. Taking $\mathrm{DM_h} \simeq 300-900\ \mathrm{pc\ cm^{-3}}$ and $t_{\mathrm{SNR}} \simeq 12-37$ years for $\alpha = 1/2$, one obtains the following constraint:

$$\left(\frac{M}{10M_\odot}\right)^{3/4} \left(\frac{E}{10^{51}\ \mathrm{erg}}\right)^{-1/4} \left(\frac{n_0}{1\ \mathrm{cm^{-3}}}\right)^{1/2} \simeq 3.5-18.6. \tag{10}$$

The SNR velocity is given by

$$v \simeq (450-1400)\ \mathrm{km\ s^{-1}} \left(\frac{E}{10^{51}\ \mathrm{erg}}\right)^{1/3} \left(\frac{n_0}{1\ \mathrm{cm^{-3}}}\right)^{1/3}. \tag{11}$$

Based on the above result of $t_{\mathrm{SNR}} \simeq (12-37)$ yr, the SNR radius is estimated to be

$$R_{\mathrm{SNR}} \simeq 0.02\ \mathrm{pc} \left(\frac{E}{10^{51}\ \mathrm{erg}}\right)^{1/3} \left(\frac{n_0}{1\ \mathrm{cm^{-3}}}\right)^{1/3}. \tag{12}$$

Since the SNR is more likely in the free-expansion phase with a typical velocity $v \gtrsim$ a few $\times\, 1000\ \mathrm{km\ s^{-1}}$, the condition of $(E/10^{51}\ \mathrm{erg})^{1/3} (n_0/1\ \mathrm{cm^{-3}})^{1/3} \gtrsim 10$ is required in this case, e.g., in a star-formation region with $n_0 \gtrsim 10^3\ \mathrm{cm^{-3}}$. Notice the above discussion assumes that the SNR DM is equal to the host DM. Considering an additional poorly-known DM contribution from the interstellar medium in the host galaxy, the above constraint would be looser.

It is worth noting that Hilmarsson et al. [44] made a constraint on the SNR properties based on the long-term RM decline of FRB 20121102A and found that the SNR has an age of $\sim$10 years. However, the potential assumption is that the field configuration always remains unchanged during the SNR expansion. For FRB 20190520B, the observed RM reversal suggests that such an assumption is invalid at least for this source. Therefore, compared with the RM variation that is modulated by the field configuration, the DM variation as the probe to constrain the SNR age can be much more robust.

On the other hand, a binary environment is also a popular model to explain the evolution of DM and RM of FRB repeaters [45–47]. However, the observed long-term DM decline of FRB 20190520B disfavors that it originates from the binary orbital motion due to the following reason. We consider that the central engine of FRB 20190520B is a neutron star with a mass of $m \sim 1.4M_\odot$, and the companion has a mass of $M_c$. According to Kepler's third law, the orbital separation is

$$\begin{aligned} a &= \left[\frac{G(M_c+m)P_{\mathrm{orb}}^2}{4\pi^2}\right]^{1/3} \\ &= 5.4\mathrm{AU} \left(\frac{M_c+m}{10M_\odot}\right)^{1/3} \left(\frac{P_{\mathrm{orb}}}{4\mathrm{yr}}\right)^{2/3}. \end{aligned} \tag{13}$$

Since FRB 20190520B has exhibited a long-term DM decay, the orbital period should satisfy $P_{\mathrm{orb}} \gg 4$ years, implying a much larger separation. The wind velocity should be at least larger than the escape velocity:

$$\begin{aligned} v_w &\gtrsim v_{\mathrm{esc}} \simeq \left(\frac{2GM_c}{R_c}\right)^{1/2} \\ &\simeq 620\ \mathrm{kms^{-1}} \left(\frac{M_c}{M_\odot}\right)^{1/2} \left(\frac{R_c}{R_\odot}\right)^{-1/2}. \end{aligned} \tag{14}$$

The electron density of the stellar wind at a distance $r$ from the companion is

$$n_w(r) \simeq \frac{\dot{M}_c}{4\pi \mu_m m_p v_w r^2}, \tag{15}$$

where $\mu_m = 1.2$ is the mean molecular weight for a solar-like composition, and $\dot{M}_c$ is the mass loss rate of the companion depending on its stellar type. We assume that the FRB emission region is close to the neutron star as its central engine. Due to $n_w \propto r^{-2}$, the companion wind would contribute most of the local DM at $r \sim a$. Thus, the DM contributed by the stellar wind is estimated as [46,48]

$$\begin{aligned} \mathrm{DM}_w &\sim n_w a \lesssim 100\ \mathrm{pc\ cm^{-3}} \left(\frac{M_c+m}{10M_\odot}\right)^{-1/3} \\ &\times \left(\frac{P_{\mathrm{orb}}}{4\mathrm{yr}}\right)^{-2/3} \left(\frac{\dot{M}_c}{10^{-6}M_\odot \mathrm{yr^{-1}}}\right) \left(\frac{v_w}{10^3\mathrm{kms^{-1}}}\right)^{-1}. \end{aligned} \tag{16}$$

Since FRB 20190520B appears to have a long-term DM decline in the past 4 years, the expected amplitude of the DM variation in one complete orbital period should be much larger than the observed value of $\Delta\mathrm{DM} \sim$ a few $\times\, 10\ \mathrm{pc\ cm^{-3}}$. In this case, the mass loss rate of the companion is constrained to be $\dot{M}_c \gtrsim 10^{-6} M_\odot \mathrm{yr^{-1}}$. Furthermore, if the main host DM with $\mathrm{DM} \sim (300-900)\ \mathrm{pc\ cm^{-3}}$ is contributed by the binary environment, the mass loss rate of the companion is required to be $\dot{M}_c \gtrsim 10^{-5} M_\odot \mathrm{yr^{-1}}$. Therefore, if the DM variation is from the binary orbital motion, it is required that the companion is at least a post-main-sequence star [46,49].

On the other hand, although the mass loss rate of the companion is still within an acceptable range, especially for a post-main-sequence star. The observed RM evolution with a significant reversal signal reported by Anna-Thomas et al. [5] does not support the binary picture. In observation, the timescale of the RM variation of FRB 20190520B is about a few months, much shorter than the timescale of the DM decline. As pointed out above, if the observed long-term DM decline is due to the orbital motion, the orbital period should be $P_{\mathrm{orb}} \gg 4$ years. In this case, monthly timescales for the RM variation should correspond to a very small orbital phase and the geometric configuration of the magnetic field along the line of sight should be relatively stable, inconsistent with the observed RM reversal of FRB 20190520B.

## 6. Discussion and conclusion

Although hundreds of FRBs have been detected, their origin remains a mystery. Some FRBs favor with core-collapse events involving compact objects like magnetars. As for FRB 20190520B, it was hypothesized that it is driven by a young magnetar, formed in the aftermath of a massive star's core collapse, situated in the overlapping structures of a magnetar wind nebula (MWN) and a supernova remnant [50]. Based on the large reversal RM evolution results, a binary model was proposed for FRB 20190520B [45].

The follow-up observations on Crab and Vela (PSR 0833–45) pulsars present a $\Delta$DM $\sim 10^{-2}$ pc cm$^{-3}$ much higher than the general pulsar $\Delta$DM from $\sim 10^{-4}$ to $10^{-3}$ pc cm$^{-3}$ [10–13]. As the electron density of the ISM is unable to fully account for the observed DM variations, the DM fluctuations in the Crab and Vela pulsars are more likely attributed to the electron density within the nebula [13,51].

The dispersion measure (DM) of FRB 20190520B exhibits a pronounced decline at a rate of $(-12.4 \pm 0.3)$ pc cm$^{-3}$yr$^{-1}$, which is significantly higher than those observed in typical pulsars or other FRBs. Consistent with previous findings, the measurements of $\tau_s$ and $\nu_d$ of FRB 20190520B favor a two-screen model, where scattering timescales are attributed to a screen near the source, while scintillation effects arise from a screen within the Milky Way. The mean measured $\tau_s$ is (11±6) ms, slightly exceeding the value reported by Ref. [41]. Compared to the decrease in DM, the scattering timescales change slightly. The new $\nu_d$ results exhibit stable values, indicating that ISM fluctuations within the Milky Way are minimal compared to the significant dynamic changes in the host galaxy. All measurements and observational results are presented in Tables S1 and S2 (online). FRB 20190520B exhibits a substantial host DM contribution, and the accompanying scattering timescale $\tau_s$ further supports that the variation originates from the local environment of the source. Assuming a power-law evolution of DM in an expanding supernova (SN) ejecta, the corresponding age of the associated SNR is estimated to be $t_{\rm SNR} \simeq 24\text{–}73\ \alpha$ yr. This implies that the SNR is likely in the free-expansion phase, with an age of approximately $10-100$ years. The large DM variation is consistent with a substantial host contribution (DM$_{\rm host}$) and supports the interpretation that the observed monotonic DM decrease arises from the expansion of a young SNR.

Our study confirms FRB 20190520B as the only known FRB exhibiting sustained activity on year-long timescales. Over approximately four years of biweekly to triweekly monitoring with FAST, bursts were consistently detected, enabling long-term investigations of the source's dispersion and scattering properties. The observed decrease in DM supports a scenario in which a magnetar is embedded within a young, expanding supernova remnant, offering valuable insights into the origin and evolution of FRB sources.

## Conflict of interest

The authors declare that they have no conflict interest.

## Acknowledgments

This work was supported by the National Natural Science Foundation of China (12203069 and 12588202). Chen-Hui Niu and Yuan-pei Yang acknowledge support from the National SKA Program of China (2022SKA0130100). Chen-Hui Niu also acknowledges support from the Basic Research Project of Central China Normal University, the Hubei QB Project, the CAS Youth Interdisciplinary Team and the Foundation of Guizhou Provincial Education Department (KY(2023)059). Yuan-Pei Yang is supported by the National Natural Science Foundation of China (12473047) and the National Key Research and Development Program of China (2024YFA1611603). Yi Feng is supported by the National Natural Science Foundation of China (12203045), the Leading Innovation and Entrepreneurship Team of Zhejiang Province of China (2023R01008), and the Key R&D Program of Zhejiang Province (2024SSYS0012).

We thank Stella Koch Ocker for her help in interpreting the scattering effect, and the FAST data center operations team for scheduling the follow-up observations and data acquisition.

## Author contributions

Chen-Hui Niu, Di Li, Yuan-Pei Yang, and Yuhao Zhu contributed to the overall conceptualization and main idea of the paper. Yuan-Pei Yang, Bing Zhang, Zexin Du, A Ming Chen, and Weiyang Wang conducted the theoretical analysis. Chen-Hui Niu, Yuhao Zhu, Yongkun Zhang, Jiaheng Zhang, Jiarui Niu, Jumei Yao, Pei Wang, Yi Feng, Chao-Wei Tsai, Chenchen Miao, and Junshuo Zhang handled the data processing for FAST observations. Yijun Hou, Chen–Hui Niu, and Wenfei Yu analyzed the potential contributions from the solar system. Chen-Hui Niu, Di Li, Bing Zhang, Weiwei Zhu, Wenfei Yu and Ji-An Jiang contributed FAST data from PI observations. Xiaoping Zheng, Xinming Li and Wenfei Yu assisted with paper writing and provided suggestive discussions.

## Appendix A. Supplementary material

Supplementary data to this article can be found online at https://doi.org/10.1016/j.scib.2025.11.023.

## References

[1] Lorimer DR, Bailes M, McLaughlin MA, et al. A bright millisecond radio nurst of extragalactic origin. Science 2007;318:777.
[2] Macquart JP, Prochaska JX, McQuinn M, et al. A census of baryons in the Universe from localized fast radio bursts. Nature 2020;581:391–5.
[3] Nan R, Li D. The five-hundred-meter aperture spherical radio telescope (FAST) project. Mater Sci Eng Conf Ser 2013;44:012022.
[4] Niu CH, Aggarwal K, Li D, et al. A repeating fast radio burst associated with a persistent radio source. Nature 2022;606:873–7.
[5] Anna-Thomas R, Connor L, Dai S, et al. Magnetic field reversal in the turbulent environment around a repeating fast radio burst. Science 2023;380:599–603.
[6] Lee KG, Khrykin IS, Simha S, et al. The FRB 20190520B sight line intersects foreground galaxy clusters. Astrophys J Lett 2023;954:L7.
[7] Feng Y, Zhang YK, Li D, et al. Circular polarization in two active repeating fast radio bursts. Sci Bull 2022;67:2398–401.
[8] Zhang YK, Li D, Feng Y, et al. The arrival time and energy of FRBs traverse the time-energy bivariate space like a Brownian motion. Sci Bull 2024;69:1020–6.
[9] Rankin JM, Roberts JA. Time variability of the dispersion of the crab nebula pulsar. In: Davies RD, Graham-Smith F, editors. The Crab Nebula. Vol. 46, IAU symposium; 1971, 114.
[10] Hamilton PA, Hall PJ, Costa ME. Changing parameters along the path to the VELA pulsar. Mon Not R Astron Soc 1985;214:5P–8.
[11] Kuzmin A, Losovsky BY, Jordan CA, et al. Correlation of the scattering and dispersion events in the Crab Nebula pulsar. Astron Astrophys 2008;483:13–4.
[12] You XP, Hobbs G, Coles WA, et al. Dispersion measure variations and their effect on precision pulsar timing. Mon Not R Astron Soc 2007;378:493–506.
[13] McKee JW, Lyne AG, Stappers BW, et al. Temporal variations in scattering and dispersion measure in the Crab Pulsar and their effect on timing precision. Mon Not R Astron Soc 2018;479:4216–24.
[14] Rankin JM, Comella JM, Craft Jr HD, et al. Radio pulse shapes, flux densities, and dispersion of pulsar NP 0532. Astrophys J 1970;162:707.
[15] Rickett BJ. Interstellar scattering and scintillation of radio waves. Ann Revf Astro and Astrophys 1977;15:479–504.
[16] Cordes JM, Weisberg JM, Boriakoff V. Small-scale electron density turbulence in the interstellar medium. Astrophys J 1985;288:221–47.
[17] Cordes JM. Space velocities of radio pulsars from interstellar scintillations. Astrophys J 1986;311:183–96.
[18] Wang N, Johnston S, Manchester RN. 13 years of timing of PSR B1259–63. Mon Not R Astron Soc 2004;351:599–606.
[19] Johnston S, Ball L, Wang N, et al. Radio observations of PSR B1259–63 through the 2004 periastron passage. Mon Not R Astron Soc 2005;358:1069–75.

[20] Phillips JA, Wolszczan A. Time variability of pulsar dispersion measures. Astrophys J Lett 1991;382:L27.
[21] Keith MJ, Coles W, Shannon RM, et al. Measurement and correction of variations in interstellar dispersion in high-precision pulsar timing. Mon Not R Astron Soc 2013;429:2161–74.
[22] Petroff E, Keith MJ, Johnston S, et al. Dispersion measure variations in a sample of 168 pulsars. Mon Not R Astron Soc 2013;435:1610–7.
[23] Demorest PB, Ferdman RD, Gonzalez ME, et al. Limits on the stochastic gravitational wave background from the north American nanohertz observatory for gravitational waves. Astrophys J 2013;762:94.
[24] Jones ML, McLaughlin MA, Lam MT, et al. The NANOGrav nine-year data set: measurement and analysis of variations in dispersion measures. Astrophys J 2017;841:125.
[25] Xu H, Niu JR, Chen P, et al. A fast radio burst source at a complex magnetized site in a barred galaxy. Nature 2022;609:685–8.
[26] Li D, Wang P, Zhu WW, et al. A bimodal burst energy distribution of a repeating fast radio burst source. Nature 2021;598:267–71.
[27] Kumar P, Luo R, Price DC, et al. Spectropolarimetric variability in the repeating fast radio burst source FRB 20180301A. Mon Not R Astron Soc 2023;526:3652–72.
[28] Metzger BD, Berger E, Margalit B. Millisecond magnetar birth connects FRB 121102 to superluminous supernovae and long-duration gamma-ray bursts. Astrophys J 2017;841:14.
[29] Jiang P, Tang NY, Hou LG, et al. The fundamental performance of FAST with 19-beam receiver at L band. Res Astron Astrophys 2020;20:064.
[30] Hotan AW, van Straten W, Manchester RN. PSRCHIVE and PSRFITS: an open approach to radio pulsar data storage and analysis. Public Astron Soc Pacific 2004;21:302–9.
[31] Barsdell BR, Bailes M, Barnes DG, et al. Accelerating incoherent dedispersion. Mon Not R Astron Soc 2012;422:379–92.
[32] Niu CH, Li D, Luo R, et al. CRAFTS for fast radio bursts: extending the dispersion-fluence relation with new FRBs detected by FAST. Astrophys J Lett 2021;909:L8.
[33] Zhang B. Fast radio burst energetics and detectability from high redshifts. Astrophys J Lett 2018;867:L21.
[34] Gourdji K, Michilli D, Spitler LG, et al. A Sample of low-energy bursts from FRB 121102. Astrophys J Lett 2019;877:L19.
[35] Seymour A, Michilli D, Pleunis Z. DM_phase: algorithm for correcting dispersion of radio signals. Astrophysics Source Code Library, record ascl: 1910.004 2019.
[36] Gupta Y, Rickett BJ, Lyne AG. Refractive interstellar scintillation in pulsar dynamic spectra. Mon Not R Astron Soc 1994;269:1035–68.
[37] Ramachandran R, Demorest P, Backer DC, et al. Interstellar plasma weather effects in long-term multifrequency timing of pulsar B1937+21. Astrophys J 2006;645:303–13.
[38] Gopinath A, Bassa CG, Pleunis Z, et al. Propagation effects at low frequencies seen in the LOFAR long-term monitoring of the periodically active FRB 20180916B. Mon Not R Astron Soc 2024;527:9872–91.
[39] Ocker SK, Cordes JM, Chatterjee S, et al. The large dispersion and scattering of FRB 20190520B are dominated by the host galaxy. Astrophys J 2022;931:87.
[40] Zhu Y, Niu C, Dai S, et al. A narrowband burst from FRB 20190520B simultaneously observed by FAST and PARKES. Chin Phys Lett 2024;41:109501.
[41] Ocker SK, Cordes JM, Chatterjee S, et al. Scattering variability detected from the circumsource medium of FRB 20190520B. Mon Not R Astron Soc 2023;519:821–30.
[42] Yang YP, Zhang B. Dispersion measure variation of repeating fast radio burst sources. Astrophys J 2017;847:22.
[43] Piro AL, Gaensler BM. The dispersion and rotation measure of supernova remnants and magnetized stellar winds: application to fast radio bursts. Astrophys J 2018;861:150.
[44] Hilmarsson GH, Michilli D, Spitler LG, et al. Rotation measure evolution of the repeating fast radio burst source FRB 121102. Astrophys J Lett 2021;908: L10.
[45] Wang FY, Zhang GQ, Dai ZG, et al. Repeating fast radio burst 20201124A originates from a magnetar/Be star binary. Nat Commun 2022;13:4382.
[46] Yang YP, Xu S, Zhang B. Faraday rotation measure variations of repeating fast radio burst sources. Mon Not R Astron Soc 2023;520:2039–54.
[47] Zhang B, Hu RC. Magnetars in binaries as the engine of actively repeating fast radio bursts. arXiv: 2508.12119, 2025.
[48] Chen AM, Guo YD, Yu YW, et al. Radio absorption in high-mass gamma-ray binaries. Astro Astrophys 2021;652:A39.
[49] Vassiliadis E, Wood PR. Evolution of low- and intermediate-mass stars to the end of the asymptotic giant branch with mass loss. Astrophys J 1993;413:641.
[50] Zhao ZY, Wang FY. FRB 190520B embedded in a magnetar wind nebula and supernova remnant: a Luminous persistent radio source, decreasing dispersion measure, and large rotation measure. Astrophys J Lett 2021;923:L17.
[51] Backer DC, Hama S, van Hook S, et al. Temporal variations of pulsar dispersion measures. Astrophys J 1993;404:636.